\documentclass[a4paper]{jpconf}
\usepackage{mkfig}
\usepackage{graphicx}
\begin{document}
\title{Implications of Recent Stellar Wind Measurements}

\author{Brian E. Wood$^1$}

\address{$^1$ Naval Research Laboratory, Space Science Division,
  Washington, DC 20375, USA}

\ead{brian.wood@nrl.navy.mil}


\begin{abstract}

     Very recent measurements of stellar winds are used
to update relations between winds and coronal activity.
New wind constraints include an upper limit of
$\dot{M}<0.1$~$\dot{M}_{\odot}$ for $\tau$~Ceti (G8~V), derived from
a nondetection of astrospheric H~I Lyman-$\alpha$ absorption.
This upper limit is reported here for the first time, and represents
the weakest wind constrained using the astrospheric absorption
technique.  A high mass loss rate measurement of
$\dot{M}=10$~$\dot{M}_{\odot}$ for $\delta$~Pav (G8~IV)
from astrospheric Lyman-$\alpha$ absorption suggests stronger winds
for subgiants than for main sequence stars of equivalent activity.
A very low mass-loss rate of $\dot{M}\approx 0.06$~$\dot{M}_{\odot}$
recently estimated for GJ~436 (M3~V) from Lyman-$\alpha$ absorption
from an evaporating exoplanetary atmosphere implies inactive M dwarfs
may have weak winds compared with GK dwarfs of similar activity.

\end{abstract}

\section{Introduction}

     The first stellar wind ever detected was naturally that of our
own Sun, by one of the earliest spacecraft \cite{mn62}.
Most stars are now known to possess
stellar winds of some sort, but hot coronal winds like that of the Sun
are among the hardest to detect.  The solar wind is quite weak
compared to the stronger and more easily detected winds of hot stars
and cool giant/supergiant stars.  There has still in fact been no
unambiguous direct detection of a coronal wind
streaming from another star.

     The primary coronal wind detection method that has arisen relies
on detecting not the wind itself, but its interaction with the
surrounding interstellar medium (ISM), i.e., the stellar astrosphere.
This technique relies on detecting H~I Lyman-$\alpha$ absorption from
the astrosphere, using high resolution UV spectra from the
{\em Hubble Space Telescope} (HST), specifically from the Goddard High
Resolution Spectrograph (GHRS) or Space Telescope Imaging Spectrograph
(STIS) instruments.  The first such detection was for the nearest star
system, $\alpha$~Cen~AB \cite{jll96}.  Numerous detections have
since followed \cite{bew05b,bew14}, but not every observed
line-of-sight (LOS) has led to a detection.  In addition to the
astrospheric absorption, absorption from our own heliosphere is also
sometimes observed in upwind directions of the ISM flow seen by the
Sun, depending on the amount of obscuring absorption from the ISM
itself.  It is worth noting that in addition to the HST Lyman-$\alpha$
studies of the astrospheres of coronal stars, the astrospheres of other
types of stars have been studied using both imaging and
spectrocopy \cite{ajvm14,hak16,bew16}.

     Mass loss rates estimated from astrospheric Lyman-$\alpha$
absorption have been used to assess how coronal winds relate to
coronal activity, as quantified by X-ray surface flux, $F_X$
\cite{bew05b,bew14}.  From these results it is possible to infer what
the solar wind was like when the Sun was younger and more active.
Constraints on wind evolution are also important for understanding how
coronal stars shed angular momentum \cite{cpj15}.  Wind effects on
exoplanets are another motivating factor for improving our
understanding of coronal winds.  Most of the exoplanets that have been
discovered orbit very close to their stars, where they will likely see
very high wind fluxes due to their close proximity
\cite{hl10,jdag16,ifs16}.

\section{New Wind Measurements}

\begin{table}[t]
\caption{Mass Loss Measurements for Coronal Winds}
\begin{center}
\begin{tabular}{lccccccc}
\br
Star & Spectral & $d$ & $V_{ISM}$ & $\theta$ & $\dot{M}$$^a$ & Log L$_{x}$ &
  Radius \\
 & Type & (pc) & (km/s) & (deg) & ($\dot{M}_{\odot}$) & &
  (R$_{\odot}$) \\
\mr
\multicolumn{8}{l}{\underline{OLD MEASUREMENTS}} \\
Prox Cen      & M5.5 Ve     & 1.30 & 25 & 79 &$<0.2$& 27.22 & 0.14 \\
$\alpha$ Cen AB&G2 V+K0 V   & 1.35 & 25 & 79 &0.46+1.54&26.99+27.32&1.22+0.86\\
$\epsilon$ Eri& K2 V        & 3.22 & 27 & 76 & 30   & 28.31 & 0.74 \\
61 Cyg A      & K5 V        & 3.48 & 86 & 46 & 0.5  & 27.03 & 0.67 \\
$\epsilon$ Ind& K5 V        & 3.63 & 68 & 64 & 0.5  & 27.39 & 0.73 \\
EV Lac        & M3.5 Ve     & 5.05 & 45 & 84 &  1   & 28.99 & 0.30\\
70 Oph AB     & K0 V+K5 V   & 5.09 & 37 &120 &55.7+44.3&28.09+27.97&0.83+0.67\\
36 Oph AB     & K1 V+K1 V   & 5.99 & 40 &134 & 8.5+6.5 &28.02+27.89&0.69+0.59\\
$\xi$ Boo AB  & G8 V+K4 V   & 6.70 & 32 &131 & 0.5+4.5 &28.91+28.08&0.86+0.61\\
61 Vir        & G5 V        & 8.53 & 51 & 98 & 0.3  & 26.87 & 0.99 \\
$\delta$~Eri  & K0 IV       & 9.04 & 37 & 41 &  4   & 27.05 & 2.58 \\
$\pi^1$ UMa   & G1.5 V      & 14.4 & 43 & 34 & 0.5  & 28.99 & 0.97 \\
$\lambda$ And & G8 IV-III   &25.8 & 53 & 89 &  5   & 30.82 & 7.40 \\
DK UMa        & G4 III-IV   & 32.4 & 43 & 32 & 0.15 & 30.36 & 4.40 \\
\multicolumn{8}{l}{\underline{NEW MEASUREMENTS}} \\
$\tau$~Ceti   & G8 V        & 3.65 & 56 & 59 &$<0.1$& 26.69 & 0.77 \\
$\delta$~Pav  & G8 IV       & 6.11 & 29 & 72 & 10   & 27.29 & 1.22 \\
GJ~892        & K3 V        & 6.55 & 49 & 60 & 0.5  & 26.85 & 0.78 \\
GJ~436        & M3 V        & 9.75 & 79 & 97 &0.059$^b$&26.76 & 0.44 \\
\br
\end{tabular}
\end{center}
$^a$Mass loss rates for individual members of binaries estimated from
  mass-loss/activity relations (see text). \\
$^b$Measured from absorption from an evaporating
  exoplanetary atmosphere instead of from astrospheric absorption.
\normalsize
\end{table}
     It has been over 20 years since the first astrospheric
detection, and HST continues to be capable of providing relevant new
data.  Nevertheless, the number of detections remains relatively
small.  Table~1 lists the detections and resulting mass-loss rate
measurements, $\dot{M}$, in solar units, with $\dot{M}_{\odot}=2\times
10^{-14}$ M$_{\odot}$~yr$^{-1}$ \cite{bew05b,bew14}.  The
$\dot{M}$ measurements require knowledge of the ISM wind vector seen
in the rest frame of the star, so Table~1 also lists the wind speed
seen by the star, $V_{ISM}$, and the angle between the upwind
direction of the ISM flow vector and our LOS to the star,
$\theta$.  The final two columns list X-ray luminosities (in
erg~s$^{-1}$ units) and stellar radii.

     Nearly all of the detections listed as ``Old Observations''
are from before 2005 \cite{bew05b}, with the exception of
$\pi^1$~UMa, which is from 2014 \cite{bew14}.  There are a
number of reasons for the sparsity of detections post-2005, and the
relatively paltry number of detections overall.  The distance column
in Table~1 provides an indication of one fundamental difficulty.
There are only three detections outside 10~pc, and two of those are
actually considered marginal \cite{bew05a}.  Two major reasons
for nondetections are high ISM column densities, meaning that ISM
Lyman-$\alpha$ absorption obscures the astrospheric absorption
signature; and a surrounding ISM that is fully ionized, meaning that
there is no astrospheric Lyman-$\alpha$ absorption at all because no
neutral H is injected into the astrosphere.  Both of these are far
more likely to occur for more distant targets, for reasons now
described.

     The Sun lies in a region of space called the Local Bubble (LB),
a region of generally low ISM density extending roughly 100~pc
from the Sun in most directions \cite{dms99,rl03}.
Most of the LB is hot and fully ionized.  However,
there are small, warm, partially neutral clouds embedded within it.
It so happens that the Sun resides within one of these, the
Local Interstellar Cloud (LIC), which is one of a number of
similar clouds in the solar vicinity \cite{sr08}.
Very nearby stars are likely to reside within either the
LIC or one of the other nearby clouds.  For such stars, there
will probably be an astrospheric absorption signature, since the
surrouding ISM is likely partly neutral, and there is also a good
chance the ISM H~I column will not be too high to obscure that
signature due to the short LOS.

     However, an undetectable astrospheric absorption signature
can also arise from a low $V_{ISM}$ value, a disadvantageous
astrosphere orientation, or a weak stellar wind.  Low
$V_{ISM}$ leads to cooler and less decelerated astrospheric H~I, which
makes the astrospheric absorption narrower and less separable from the
ISM absorption.  It is not a coincidence that all of the detections
in Table~1 have $V_{ISM}$ higher or comparable to the solar value
of $V_{ISM}=26.08\pm 0.21$ km~s$^{-1}$ \cite{bew15}.  As for
astrosphere orientation, it is easier to detect
astrospheres in upwind directions (i.e., $\theta<90^{\circ}$).
All but three of the detections in Table~1 have $\theta<100^{\circ}$.
As for the stellar wind, a weak wind naturally means a smaller
astrosphere with lower H~I column densities, and therefore weaker
and narrower astrospheric absorption.  Measurements of mass-loss
rates from astrospheric absorption rely on this effect.

     The low detection fraction of astrospheres, particularly at
distances over 10~pc, makes it very difficult to obtain HST observing
time explicitly to detect more astrospheres.  Most of the
detections in Table~1 are actually based on archival HST spectra
taken for reasons having nothing to do with astrospheres, or even
with the H~I Lyman-$\alpha$ line.  And this leads to the explanation
for why there have been so few detections since 2005.  The golden age
for this kind of research was 1997--2004, after the STIS instrument
was installed on HST (replacing GHRS), but before STIS suffered a
failure in 2004.  During this period, STIS was the primary UV
spectrometer on HST, and many observers used its E140M grating to
observe the entire 1150--1700~\AA\ spectral region for various
reasons, most having nothing to do with the H~I Lyman-$\alpha$ line at
1216~\AA.  This filled the HST archives with spectra that could be
used to search for astrospheric absorption.  Except for $\alpha$~Cen
and $\pi^1$~UMa, all of the ``Old Measurements'' of astrospheres
listed in Table~1 are from data taken during this time
period \cite{bew05a}.

     The STIS instrument was repaired during the last HST servicing
mission in 2009, but this mission also installed another UV
spectrometer onto HST, the Cosmic Origins Spectrograph (COS).
The newer COS has much higher sensitivity than STIS and is therefore
now the dominant FUV spectrometer.  However, COS observations cannot be
used for astrospheric studies, because COS has insufficient spectral
resolution and its large apertures let in too much geocoronal
Lyman-$\alpha$ emission.  Thus, since 2004 the HST archive has not
been accumulating high-resolution Lyman-$\alpha$ spectra usable for
astrospheric research at the rate it was in 1997--2004.

     Nevertheless, there are two very recent astrosphere detections
that have been reported just in the past year, for $\delta$~Pav (G8~IV)
and GJ~892 (K3~V), as listed in Table~1.  For $\delta$~Pav, a
mass-loss rate of $\dot{M}=10~\dot{M}_{\odot}$ is inferred
\cite{jz18}, while for GJ~892 the measurement is
$\dot{M}=0.5~\dot{M}_{\odot}$ \cite{cpf18}.
It is noteworthy that $\delta$~Pav and GJ~892 are both within 7~pc.
For targets within 7~pc, there are now 13 independent lines of sight
with HST Lyman-$\alpha$ spectra possessing sufficient spectral resolution
for a definitive Lyman-$\alpha$ search for astrospheric absorption.
(Proxima~Cen is a distant companion of $\alpha$~Cen, and so is not
considered an independent LOS from $\alpha$~Cen.)  Ten of
the 13 lines of sight within 7~pc lead to successful detections of
astrospheric absorption.

     This 76\% detection fraction means that the ISM within 7~pc must
be nearly or entirely filled with partially neutral ISM material.
This has the important ramification that if all stars
within 7~pc are surrounded by partially neutral ISM, then meaningful
upper limits for $\dot{M}$ can in principle be measured for the
nondetections of astrospheric absorption.  The primary reason that
upper limits are not generally estimated from nondetections is that a
likely reason for most of the nondetections (especially beyond 10~pc)
is that the surrounding ISM is fully ionized, meaning that there would
be no astrospheric Lyman-$\alpha$ absorption regardless of wind
strength.  A meaningful upper limit can only be made if this
explanation can be excluded, as it could for Proxima~Cen given that
astrospheric absorption is detected for Proxima~Cen's companion,
$\alpha$~Cen~AB.  The only previous $\dot{M}$ upper limit inferred
from a nondetection is for Proxima~Cen (see Table~1).

     The three other nondetections within 7~pc are $\tau$~Ceti (G8~V),
40~Eri~A (K1~V), and AD~Leo (M3.5~Ve) \cite{bew05a}.  Based on
the preceding paragraph, meaningful upper limits for $\dot{M}$ can in
principle be made from the nondetections of astrospheric absorption
in the Lyman-$\alpha$ spectra.  However, in practice I exclude 40~Eri~A
and AD~Leo, for very different reasons.  For AD~Leo, the ISM flow
speed is an unfavorably low $V_{ISM}=13$ km~s$^{-1}$ value.  When
combined with the surprisingly high ISM column (in cm$^{-2}$) of $\log
N_H=18.47$ for this LOS, which is the highest yet observed within
15~pc, I conclude that any $\dot{M}$ upper limit that might be derived
for AD~Leo would be too high to be useful.

     For 40~Eri~A the problem is an extremely high ISM wind velocity
of $V_{ISM}=127$ km~s$^{-1}$.  It was emphasized above how a low
$V_{ISM}$ speed is bad for astrosphere detectability, but it turns
out that an extremely high $V_{ISM}$ value can be a problem as well,
as it can make astrospheres so small and the astrospheric H~I so hot
that H~I Lyman-$\alpha$ becomes optically thin \cite{bew03}.
This greatly complicates the task of inferring constraints on the wind
from the absorption, so I simply ignore 40~Eri~A for now.
(See \cite{bew03} for more details.)

     This leaves only $\tau$~Ceti, which has very advantageous
values of $V_{ISM}=56$ km~s$^{-1}$ and $\theta=59^{\circ}$, and a
low ISM column density of $\log N_H=18.01$.  Furthermore, $\tau$~Ceti
is particularly close, with $d=3.65$~pc, making it even more likely to
be surrounded by partially neutral ISM material like that around the
Sun.  Thus, an $\dot{M}$ upper limit for $\tau$~Ceti is inferred here
for the first time.  This requires the assistance of hydrodynamical
models of the astrosphere.  However, rather than compute new models,
which can be time-consuming, existing models for the 61~Vir
astrosphere are used instead.  These are relevant thanks to the
similar $V_{ISM}=51$ km~s$^{-1}$ value of 61~Vir compared to
$\tau$~Ceti's $V_{ISM}=56$ km~s$^{-1}$.  Note that the recent
$\delta$~Pav and GJ~892 $\dot{M}$ measurements were similarly made
using existing astrospheric models of other stars as proxies \cite{jz18}.

\begin{figure}[t]
\plotfiddle{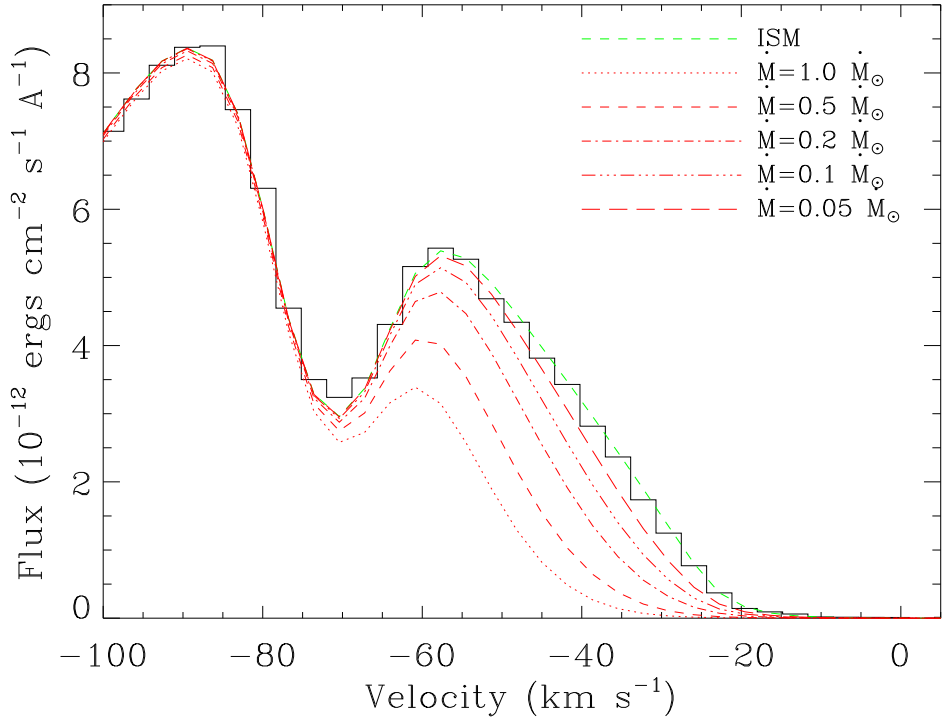}{3.5in}{90}{60}{60}{220}{-50}
\caption{A close-up of the blue side of the H~I Lyman-$\alpha$
  absorption observed towards $\tau$~Ceti, plotted on a heliocentric
  velocity scale.  The narrow absorption at $-70$ km~s$^{-1}$ is
  from interstellar deuterium.  The green dashed line shows the
  ISM absorption, which provides a reasonably good fit to the
  data.  Red lines show additional absorption predicted by
  astrospheric models assuming different mass-loss rates of
  $\dot{M}=0.05-1.0~\dot{M}_{\odot}$.  The $\dot{M}=0.1~\dot{M}_{\odot}$
  model is assumed to represent an $\dot{M}$ upper limit for
  $\tau$~Ceti.}
\end{figure}
     The full Lyman-$\alpha$ spectrum and ISM absorption fit to
the $\tau$~Ceti data can be found in \cite{bew05a}.  Figure~1 zooms
into the blue side of the H~I Lyman-$\alpha$ absorption profile where the
astrospheric absorption would be observed.  This is an astrospheric
nondetection, meaning that the absorption could be reasonably well fit
with ISM absorption alone.  Predicted astrospheric absorption is shown
for five values of $\dot{M}$, from $\dot{M}=0.05-1.0~\dot{M}_{\odot}$.
Models actually only exist for $\dot{M}\geq 0.2~\dot{M}_{\odot}$, so
the $\dot{M}=0.05~\dot{M}_{\odot}$ and $\dot{M}=0.1~\dot{M}_{\odot}$
models are actually made by extrapolation from the higher $\dot{M}$
models.  It is obvious that a $\dot{M}=0.5-1.0~\dot{M}_{\odot}$ wind
should have yielded easily detectable absorption towards $\tau$~Ceti,
but detectability naturally decreases for lower $\dot{M}$ values.  It
is dubious whether the $\dot{M}=0.05~\dot{M}_{\odot}$ model predicts a
sufficient amount of absorption to be considered detectable.  Relying
mostly on subjective judgment, I conclude that it does not and settle
on a more conservative $\dot{M}<0.1~\dot{M}_{\odot}$ upper limit for
$\tau$~Ceti, which is what is reported in Table~1.  This is the
weakest wind ever measured using the astrospheric absorption
technique, slightly lower than the $\dot{M}=0.15~\dot{M}_{\odot}$
value for the marginal DK~UMa detection \cite{bew05b}.

     There is one final new $\dot{M}$ measurement listed in
Table~1 that has not yet been mentioned, for the transiting exoplanet
host star GJ~436 (M3~V).  For this star, Lyman-$\alpha$ absorption has
been observed during exoplanet transits, implying an evaporating
planetary atmosphere, and the amount of absorption has been used to
infer the strength of the stellar wind \cite{aav17}.
Thus, this $\dot{M}=0.059^{+0.074}_{-0.040}~\dot{M}_{\odot}$
constraint is listed in Table~1 as well, though the measurement is
from exoplanetary instead of astrospheric Lyman-$\alpha$ absorption.

\section{Revised Wind-Activity Relation}

     It is natural to look for correlations between wind properties
and the properties of the stellar coronae that produce the winds.
The most readily available coronal property for comparison is
the coronal X-ray luminosity, so values of $\log L_X$ are
provided in Table~1.  Many of these are from ROSAT PSPC
measurements \cite{js04}, but newer {\em Chandra}
measurements are used when available \cite{bew18}.

\begin{figure}[t]
\plotfiddle{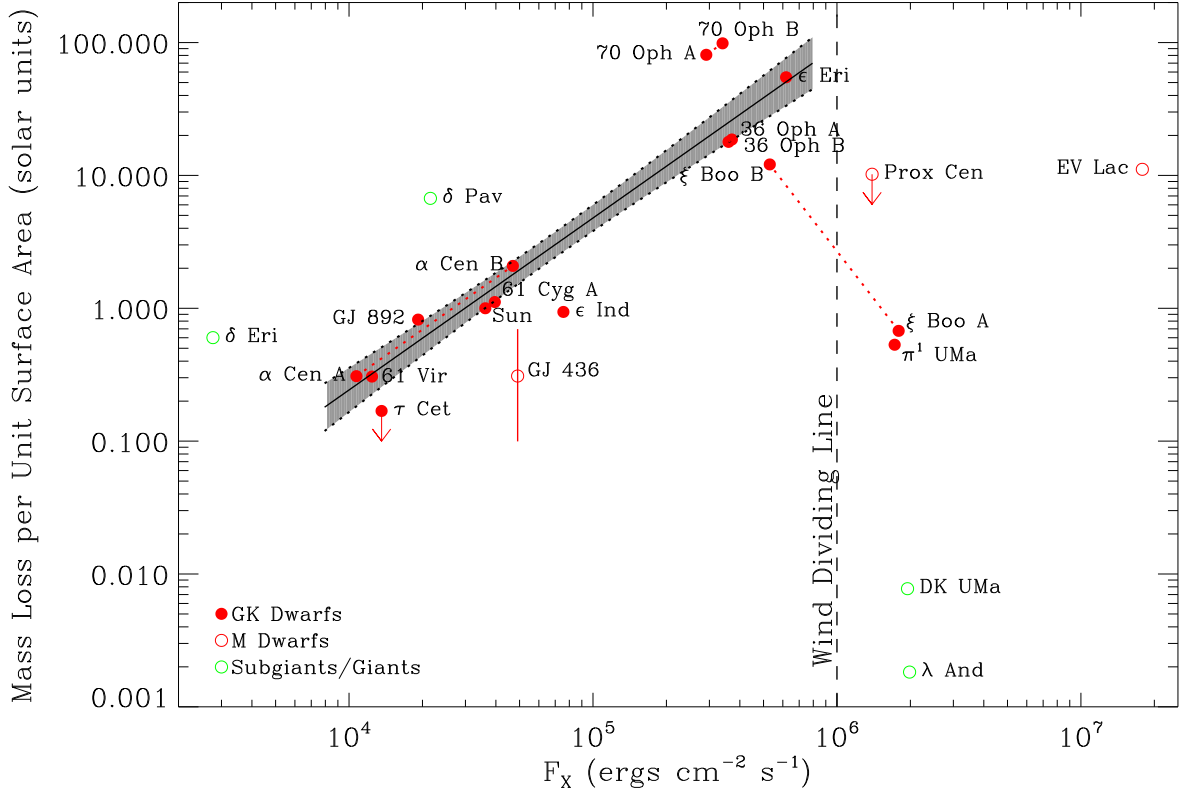}{3.9in}{0}{90}{90}{-270}{-340}
\caption{A plot of mass loss rate (per unit surface area) versus X-ray
  surface flux for all main sequence stars with measured winds,
  analogous to previously published figures \cite{bew05b,bew14}.
  Different symbols are used to represent GK dwarfs, M dwarfs,
  and subgiant/giant stars.  A power law,
  $\dot{M}\propto F_X^{1.29\pm 0.16}$, is fitted to the GK dwarfs
  where a wind/X-ray relation seems to exist, but this relation seems
  to fail for stars to the right of the ``Wind Dividing Line.''
  Separate points are plotted for the two members of four binary
  systems, where for the three less active cases the power law
  relation is used to estimate the wind contributions of 
  individual stars.  In order to yield consistency with stars of
  similar $F_X$, for the $\xi$~Boo binary $\xi$~Boo~B is
  assumed to account for 90\% of the wind and $\xi$~Boo~A
  only 10\%.}
\end{figure}
     Figure~2 plots mass loss rate per unit surface area versus X-ray
surface flux for the stars in Table~1.  A correlation is seen
for GK dwarfs with $\log F_X<6.0$, and a power law relation is
fitted to the data.  This relation does not extend beyond the
$\log F_X=6.0$ ``Wind Dividing Line,'' above which
surprisingly weak winds are observed.  This figure is very much
analogous to similar figures that have been published in the
past, including the power law fit and wind dividing
line \cite{bew05b,bew14}.  However, the figure is updated in a
number of ways.  The most obvious is the inclusion of the four new
measurements listed in Table~1.  There have also been relatively
minor revisions to the X-ray luminosities and stellar radii.
The resulting change to the power law relation is insignificant:
$\dot{M}\propto F_{x}^{1.29\pm 0.16}$ compared to the previous
$\dot{M}\propto F_{x}^{1.34\pm 0.18}$ \cite{bew05b}.  The
new GJ~892 measurements is completely consistent with the
relation.  However, the new upper limit for $\tau$~Ceti looks
anomalously low.  The significance of this inconsistency is marginal,
as the $\epsilon$~Ind data point is a similar distance below the
power law relation.

     Another difference from previous studies lies in our treatment
of binary systems.  There are four systems listed in Table~1
where both members of the binary lie within the same
astrosphere ($\alpha$~Cen~AB, 70~Oph~AB, 36~Oph~AB, and $\xi$~Boo~AB),
meaning both stellar winds will be contributing to the measured
mass-loss rate.  For the three binaries that lie entirely to the left
of the wind dividing line, the power law relation can be used to
estimate the contributions of each individual star to the binary's
collective mass-loss rate, and the results are indicated in both
Table~1 and Figure~2.  Separate X-ray luminosities are listed for
the two stars as well.  The most interesting of these three cases
is $\alpha$~Cen~AB, where the system's wind is predicted to be
dominated by the secondary star, $\alpha$~Cen~B, with
$\dot{M}=1.54~\dot{M}_{\odot}$ compared to $\dot{M}=0.46~\dot{M}_{\odot}$
for $\alpha$~Cen~A.  The $\xi$~Boo~AB system is more complicated,
with the primary to the left of the dividing line and the
secondary to the right.  In order for the two stars to be consistent
with other stars of similar $F_X$, $\xi$~Boo's wind must be
dominated by the secondary.  Consistent with past work, 
a division with $\dot{M}=4.5~\dot{M}_{\odot}$ for $\xi$~Boo~B and
$\dot{M}=0.5~\dot{M}_{\odot}$ for $\xi$~Boo~A is assumed \cite{bew10}.

     The new $\delta$~Pav measurement has interesting implications
for the winds of coronal subgiants/giants.  Both $\delta$~Pav and
$\delta$~Eri imply that inactive subgiants have significantly
stronger winds per unit surface area compared to main sequence stars
of similar $F_X$.  This is not a surprising result given that subgiants
have lower surface gravities and surface escape speeds
than main sequence stars, which should make it easier for coronal
material to escape \cite{jz18}.  In contrast,
the very active subgiant/giant stars DK~UMa and $\lambda$~And
seem to have remarkably weak winds, perhaps implying that these
stars have strong, global fields that inhibit mass loss in some way.

     The GJ~436 measurement in Figure~2 provides a first
indication that the winds of inactive M dwarfs may be weaker
per unit surface area than GK dwarfs with similar $F_X$.  Obviously,
additional measurements are required to support this conclusion.
Fortunately, these measurements should be forthcoming,
as a proposed  HST Lyman-$\alpha$ survey of 10 M dwarfs within 7~pc
has recently been accepted, and these observations should be carried
out in the coming year.  In the near future, these data will
hopefully allow the M dwarf $\dot{M}-F_X$ relation to be
characterized as well as the GK dwarf relation in Figure~2.

\ack

Support for this project was provided by NASA award NNH16AC40I to the
Naval Research Laboratory.

\section*{References}



\begin{thebibliography}{99}
\bibitem{mn62}
Neugebauer M, and Snyder C W 1962 {\it Science} {\bf 138} 1095
\bibitem{jll96}
Linsky J L, and Wood B E 1996 {\it ApJ} {\bf 463} 254
\bibitem{bew05b}
Wood B E, M\"{u}ller H -R, Zank G P, Linsky J L, and Redfield S 2005
  {\it ApJ} {\bf 628} L143
\bibitem{bew14}
Wood B E, M\"{u}ller H -R, Redfield S, and Edelman E 2014
  {\it ApJ} {\bf 781} L33
\bibitem{ajvm14}
van Marle A J, Decin L, and Meliani Z 2014 {\it A\&A} {\bf 561} A152
\bibitem{hak16}
Kobulnicky H A {\it et al} 2016 {\it ApJS} {\bf 227} 18
\bibitem{bew16}
Wood B E, M\"{u}ller H -R, and Harper G M 2016 {\it ApJ} {\bf 829} 74
\bibitem{cpj15}
Johnstone C P, Zhilkin A, Pilat-Lohinger E, Bisikalo D, G\"{u}del M, and
  L\"{u}ftinger T 2015 {\it A\&A} {\bf 577} A28
\bibitem{hl10}
Lammer H {\it et al} 2010 {\it Astrobiology} {\bf 10} 45
\bibitem{jdag16}
Alvarado-G\'{o}mez J D {\it et al} 2016 {\it A\&A} {\bf 594} A95
\bibitem{ifs16}
Shaikhislamov I F {\it et al} 2016 {\it ApJ} {\bf 832} 173
\bibitem{bew05a}
Wood B E, Redfield S, Linsky J L, M\"{u}ller H -R, and Zank G P 2005
  {\it ApJS} {\bf 159} 118
\bibitem{dms99}
Sfeir D M, Lallement R, Crifo F, and Welsh B Y 1999 {\it A\&A} {\bf 346} 785
\bibitem{rl03}
Lallement R, Welsh B Y, Vergely J L, Crifo F, and Sfeir D M 2003
  {\it A\&A} {\bf 411} 447
\bibitem{sr08}
Redfield S, and Linsky J L 2008 {\it ApJ} {\bf 673} 283
\bibitem{bew15}
Wood B E, M\"{u}ller, H -R, and Witte M 2015 {\it ApJ} {\bf 801} 62
\bibitem{jz18}
Zachary J, Redfield S, Linsky J L, and Wood B E 2018 {\it ApJ} {\bf 859} 42
\bibitem{cpf18}
Folsom C P {\it et al} 2018 {\it MNRAS} in press
\bibitem{bew03}
Wood B E, Linsky J L, M\"{u}ller H -R, and Zank G P 2003
  {\it ApJ} {\bf 591} 1210
\bibitem{aav17}
Vidotto A A, and Bourrier V 2017 {\it MNRAS} {\bf 470} 4026
\bibitem{js04}
Schmitt J H M H, and Liefke C 2004 {\it A\&A} {\bf 417} 651
\bibitem{bew18}
Wood B E, Laming J M, Warren H P, and Poppenhaeger K 2018 {\it ApJ} in press
\bibitem{bew10}
Wood B E, and Linsky J L 2010 {\it ApJ} {\bf 717} 1279

\end{thebibliography}
\end{document}